\documentclass[twocolumn]{aastex62}
\usepackage{graphicx,amsmath,bm,natbib,color}

\begin{document}

\bibliographystyle{apj}
\title{Kinematics of cluster galaxies and their relation to galaxy evolution} 
\author{Susmita Adhikari}
\affil{Kavli Institute for Particle Astrophysics and Cosmology, Stanford University, 452 Lomita Mall Stanford, CA 94305, USA}
\affil{Department of Astronomy, University of Illinois at Urbana-Champaign, 1002 W. Green Street, Urbana, IL 61801, USA}
\author{Neal Dalal}
\affil{Perimeter Institute for Theoretical Physics, 31 Caroline St N, Waterloo, ON N2L 2Y5, Canada }
\author{Surhud More}
\affil{Kavli Institute for the Physics and Mathematics of the Universe (WPI), University of Tokyo, 5-1-5 Kashiwanoha, Kashiwa-shi, Chiba, 277-8583, Japan}
\author{Andrew Wetzel}
\affil{Department of Physics, University of California Davis,  Davis, CA 95616, USA}

\begin{abstract}
We study the kinematics of galaxies within massive clusters, as a probe of the physics of star-formation quenching within clusters.  Using N-body simulations, we argue that satellite kinematics provide information about galaxy infall that is complementary to the (instantaneous) spatial distribution of satellites.  Comparing the simulation results with measurements of real cluster galaxies, we find evidence that the kinematics of red (quiescent) satellite galaxies are consistent with earlier infall times than that of blue (star-forming) satellites.

\end{abstract}

\keywords{galaxy evolution -- clusters --{\tiny } simulation}

\section{Introduction} \label{intro} 

A striking characteristic of observed galaxy clusters is that the fraction of red, elliptical galaxies with little or no ongoing star formation is significantly higher within clusters than in the field, and furthermore the quenched fraction increases towards small radii within clusters \citep{Balogh:1997bw,Poggianti:1999xh, Dressler:1980wq, Dressler:1984kh, Oemler:1974yw}. There are primarily two broad classes of theories that attempt to explain this behavior. 
The first class of theories invokes quenching of star formation due to astrophysical processes internal to clusters, due to a variety of 
processes like strangulation \citep{Larson:1980mv}, harassment, ram-pressure stripping \citep{Abadi:1999qy,GunnGott72}, or tidal disruption.
In this scenario, the quenching process initiates when galaxies enter their hosts, although note that quenching may require many gigayears to complete \citep{Wetzel:2012nn}.  

Another possible scenario to explain the statistics of quenched galaxies without invoking any intra-cluster processes was proposed by 
\citet{Hearin:2013foa}.  They showed that simple age-matching models can reproduce the observed 2-point statistics of both red and blue galaxies.  Unlike the quenching models described above, age-matching models explain quenched fractions not using astrophysical processes that are triggered on entry into a cluster, but instead using processes internal to galaxies that are related solely to the mass growth of its parent halo. If the supply of gas into a halo tracks the dark matter accretion rate \citep{Wetzel:2014hqa,Feldman2016}, then the ages of the stellar population within galaxies may be expected to correlate with halo formation time.
Since subhalos living in large-scale overdensities form earlier than comparable field halos, age-matching can thus naturally explain why satellite galaxies have older (redder) stellar populations than field galaxies.

Since there are multiple plausible models to account for the environmental dependence of quiescent fractions, there is motivation to look for new signatures that will help distinguish between these scenarios.  The recent detection of the splashback feature \citep{More2016,Baxter:2017csy} also adds a new dimension to this picture. The splashback feature is a density caustic, associated with the first apocentric passage of  particles or subhalos after they begin to orbit inside their host halos \citep{Adhikari2014,Diemer2014,Shi:2016lwp}. The splashback radius forms the boundary between the ``virialized" (or multi-streaming) region and the infall (single-stream) region of the halo, and it can be located far beyond the nominal ``virial'' radius of a halo defined by the virial overdensity $\Delta_{\rm vir}$ depending on the growth history of the halo \citep{More2015, Adhikari2014, Wetzel:2014hqa}.  For example, \citet{Wetzel2014} showed that the enhanced quiescent fraction of central galaxies near  groups and clusters that extends out to several virial radii $r_{\rm vir}$  can be explained by treating them like normal satellites that have had one passage through the halo (backsplash halos). These galaxies are outside the nominal virial radius but are in fact on orbits in the halo potential, within the splashback radius which can be much larger than $r_{\rm vir}$ depending on the accretion rate of the host and also the mass of the satellite. 
In fact, 
\citet{Baxter:2017csy} quantify the red fraction around SDSS RedMaPPer galaxy clusters, and show that the red fraction changes sharply at exactly the splashback radius of the halo, approaching the background value almost immediately beyond splashback.  That result does not necessarily favor one quenching scenario over another, since 
this behavior is expected in both classes of quenching models discussed above.  A related result that is perhaps more revealing is that a significant splashback feature is found using either red galaxies only \citep{Baxter:2017csy} or blue galaxies only \citep{Zuercher_inprep}.  The latter result shows that at least some satellites can remain blue (i.e., unquenched) after a full orbit within a cluster, corresponding to several gigayears.  This result is natural within age-matching models, whereas models invoking intra-cluster quenching would be ruled out if they predict fast ($\lesssim$ Gyr) quenching for all satellites.

\citet{Chamberlain2014} proposed an alternative method to probe the ages of satellites in their host.  Using physical arguments validated by N-body simulations, they showed that subhalos become spatially uncorrelated with each other beyond their tidal radii after a few dynamical times within their hosts.  Therefore, subhalo spatial correlations can be used as a clock, indicating the time since infall into their hosts. \citet{Fang2016} measured this signal using satellites in RedMaPPer clusters. They found that a signicant fraction of the reddest galaxies have fallen in recently, arguing in favor of galaxies being quenched even prior to their infall into the host.  One possible explanation for this behavior is that 
a fraction of galaxies have entered their host clusters as members of subgroups, and were quenched before infall into their current hosts \citep{Zabludoff:1998nx, McGee:2009cc, Wetzel:2012nn}.  This should lead to an {\em enhanced} quiescent fraction in galaxies that are parts of subgroups within the cluster environment, a prediction that can be easily tested by repeating the analysis of \citet{Fang2016} for blue satellites, and comparing to the previous result for red satellites.

The studies mentioned above have concentrated mostly on the spatial distribution of quenched satellites to probe the physics responsible for their enhanced densities in clusters. The velocity distribution of satellite galaxies provides complementary information to the spatial distribution, and together they encode the dynamical history of the satellite population.  This was studied in detail by \citet{Oman:2013wf}, who analyzed the orbits of simulated subhalos to construct probability distributions of infall times at each point in projected phase space.  Subsequently, \citet{Oman2016} applied these results to galaxies in real clusters, and argued that essentially {\em all} galaxies are quenched at first infall, within $3-5$ Gyrs of their entry within $2.5~r_{\rm vir}$ of the host.  This would correspond to quenching by the time of first pericentric passage, long before satellites splash back to their first apocentric passage.  This conclusion appears to be inconsistent with the detection of a splashback feature in blue galaxies in clusters \cite{Zuercher_inprep}, which (if correct) would demonstrate that not all galaxies quench within even several Gyr of entering $r_{\rm vir}$, let alone the much longer time needed to cross 2.5$r_{\rm vir}$.

This apparent inconsistency is somewhat puzzling, since the basic approach proposed by \citet{Oman:2013wf} should be valid: satellites' kinematics are certainly related to their histories, which therefore should shed light on the processes important in quenching of satellites.  For example, if intracluster physics is responsible for galaxy quenching then the effect can be significantly stronger for objects on plunging orbits, that reach far inside the halo, compared to those on orbits with higher angular momentum.  Satellites that are on grazing orbits may therefore be less affected by tidal stripping, ram pressure-stripping, etc., which are all stronger near the cluster center, leading to those orbits being populated by a higher proportion of star-forming galaxies, while plunging orbits would have a higher proportion of quenched galaxies.  Conversely, if intracluster physics is unimportant in quenching of satellite galaxies, that we might expect to see little dependence of color on orbital kinematics at a given radius.  Motivated by these conflicting results, in this paper we revisit the use of kinematics to understand galaxy evolution in clusters.  We will use an approach similar in spirit to \citet{Oman2016}, though with a slightly different goal: we will use kinematics to try to distinguish between different scenarios for quenching, namely those involving intracluster physics like strangulation etc., and those that do not, like age-matching, rather than trying to determine quenching times within the context of environmental quenching models.

\begin{figure}
	\includegraphics[width=1\linewidth, trim= 0.5in 0.0in 0in 0in,clip]{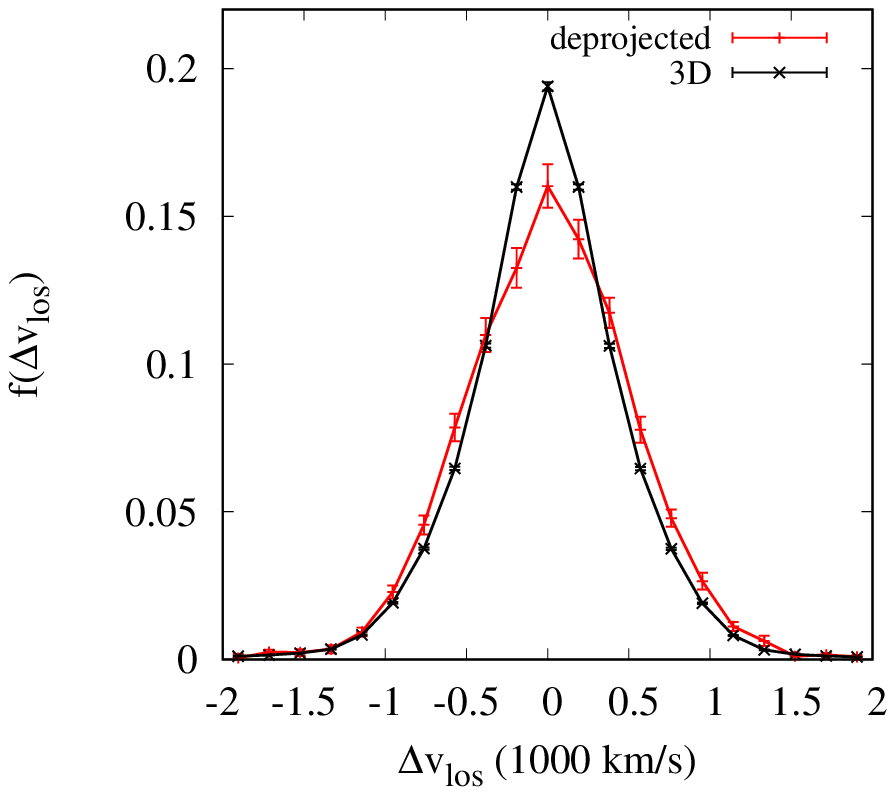}
	\includegraphics[width=0.95\linewidth, trim= 0.5in 0.0in 0.3in 0in,clip]{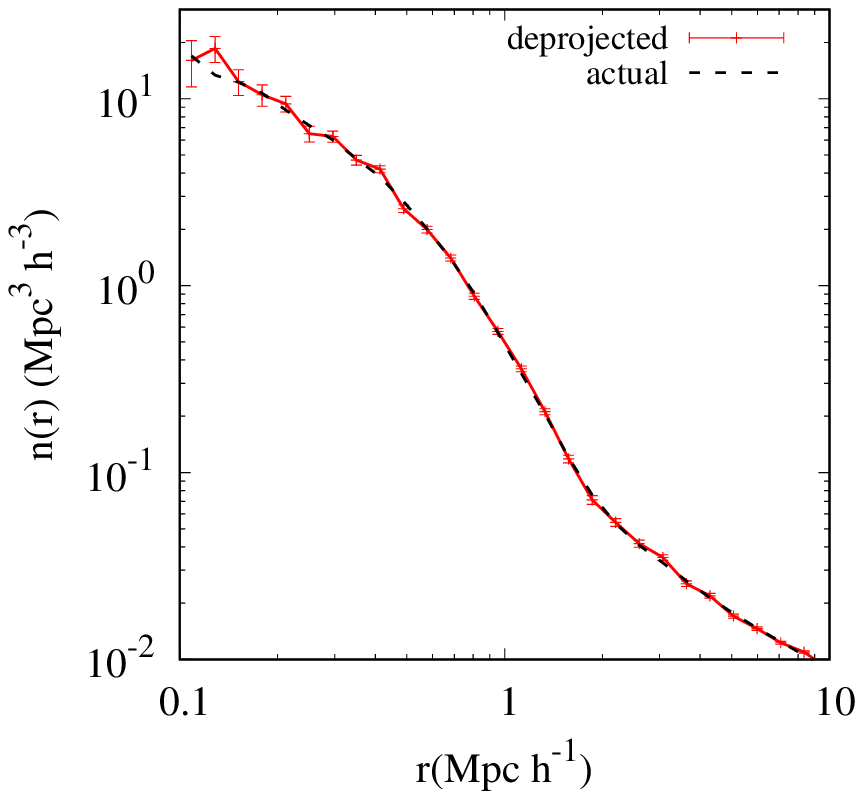}
	\caption{MDPL2 Simulations: Comparison between deprojected histogram and actual 3D distribution of subhalos between 0.6 and 0.9 Mpc $h^{-1}$ around cluster mass halos of $1-2\times 10^{14} M_\odot h^{-1}$. The top panel shows the LOS velocity histograms and the bottom panel shows the number density profiles.\label{fig:deproject}}
\end{figure}

\begin{figure*}	
	\centering
	\includegraphics[width=0.90\linewidth, trim= 0.05in 0.6in 0.05in 0.5in,clip]{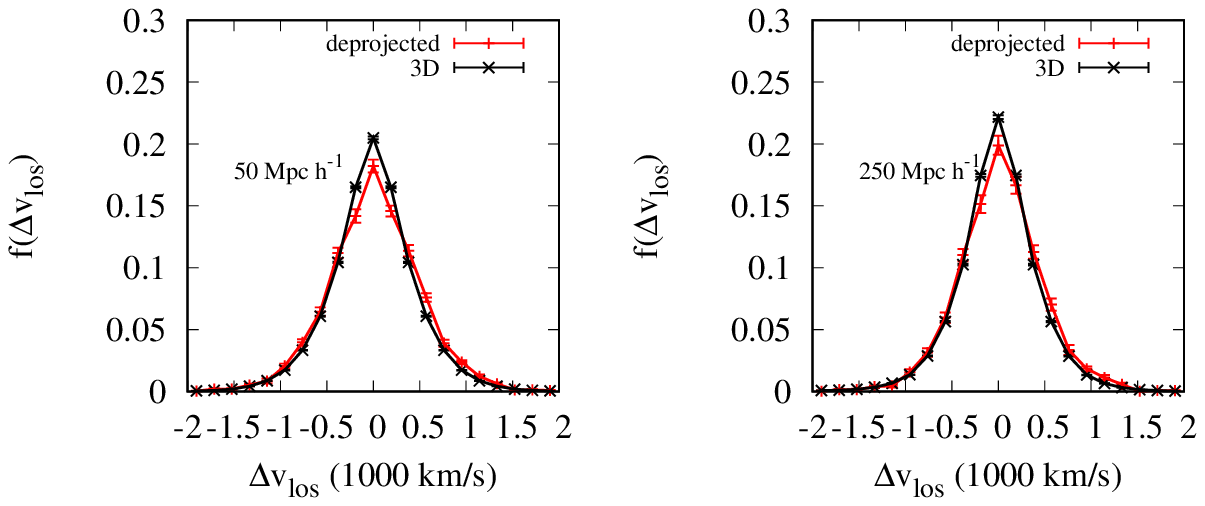}
	\includegraphics[width=0.85\linewidth, trim= 0.0in 0.5in 0.15in 0.7in,clip]{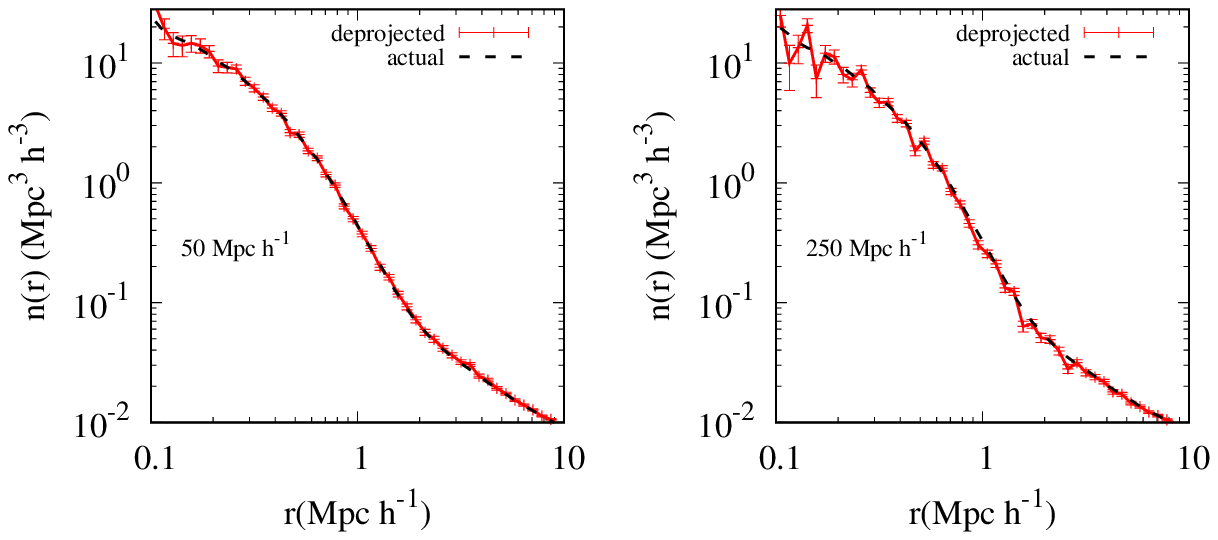}
	\caption{MDPL2 simulations: (Top) Comparison between deprojected histogram of LOS velocities and actual 3D distribution of LOS velocities of subhalos between 0.6 and 0.9 Mpc $h^{-1}$ around clusters stacked based on 2D richness, 20 $<\Lambda<$ 40. (Bottom) The same comparisons for the number density profiles. The left panels have a projection depth of $50$ Mpc $h^{-1}$ while the right panel has a projection depth of $250$ Mpc$\,h^{-1}$.\label{fig:anisotropy}}
\end{figure*}

\section{Simulation results}
\label{cdmsdm}

To understand the  dynamics of galaxies inside clusters we first study the dynamics of subhalos in $\Lambda$CDM N-Body simulations. We utilize subhalo catalogs created from the publicly available MDPL2 simulation \citep{MultiDark2}, which contains 
$3840^3$ particles in a ($1\,h^{-1}$ Gpc)$^3$ volume with a particle mass of $m_p=1.5\times10^9 M_\odot h^{-1}$.  We use the Rockstar catalogs of halos and subhalos \citep{Rockstar} publicly available at \texttt{cosmosim.org}.  We mainly look at the distribution of structure around cluster mass hosts with $M_{\rm vir} > 10^{14} M_\odot h^{-1}$ at $z=0$.  We do not distinguish between halos and subhalos when finding neighbors, since backsplash halos may be found beyond the nominal virial radius of the host \citep{More2015, Wetzel:2013sja,Adhikari2014,Diemer2014}.  This host mass range approximately matches the masses for the real clusters we discuss below in section \ref{sec:data}.
For our analysis we have only considered subhalos with peak velocities $v_{\rm peak}>200$ km/s, corresponding to massive subhalos that are comparable to the bright galaxies targeted spectroscopically in the SDSS DR7 survey.

Numerical simulations provide full 6-dimensional phase space data for all objects, but observationally we can measure only line-of-sight velocities and projected separations between galaxies in cluster fields.  This complicates the interpretation of kinematics measurements in clusters, since (for example) interlopers can contaminate observed samples.  
To isolate the cluster-specific effects, we would ideally like to remove interloper contamination.  In individual clusters, this is generally not possible, but in large ensembles of clusters, we can isolate cluster effects statistically.  
Assuming that the sample of cluster halos is viewed isotropically on average, we can use standard deprojection methods like the Abel integral transform \citep{BinneyTremaine} to reconstruct the 3D distribution of cluster galaxies using 2D projected measurements \citep{Eisenstein:2002sx}.  As an example, suppose that we measure the average (stacked) surface density distribution $\Sigma(R)$ of galaxies as a function of projected separation $R$ around cluster centers, assuming circular symmetry.  Then the average radial distribution of galaxies $\rho(r)$, as a function of 3D radius $r$, is given by  
\begin{equation}
\rho(r)=-\frac{1}{\pi}\int_{r}^{\infty}\frac{d\Sigma(R)}{dR}\frac{1}{\sqrt{R^2-r^2}}dR.
\label{eqn:rho}
\end{equation} 
Writing the projected surface density as
\begin{equation}
\Sigma(R)=\sum_{i}\frac{\delta(R-R_i)}{2\pi R_i},
\label{eqn:sigma}
\end{equation}  
where $R_i$ is the projected radius of galaxy $i$, then 
the total number of galaxies in a 3D radial bin extending from $r=a$ to $r=b$ is
\begin{eqnarray}
\label{eqn:recon}
\Delta N&=&4\pi\int_{a}^{b}\rho(r)r^2 dr\\ \nonumber
&=&\sum_{i}
\begin{cases}
g(R_i/a)-g(R_i/b) & R_i>b\\
g(R_i/a)+1 & a < R_i<b\\
0 & R_i < a
\end{cases} ,
\end{eqnarray}
where 
\begin{equation}
g(x)=\frac{2}{\pi}\left[\frac{1}{\sqrt{x^2-1}}-\tan^{-1}\left(\frac{1}{\sqrt{x^2-1}}\right)\right].
\label{eqn:delta2}
\end{equation}
The lower panel of Fig.\ \ref{fig:deproject} illustrates that deprojection correctly reconstructs the average space density of subhalos around cluster-sized halos of mass 
$M=(1-2)\times 10^{14} M_\odot h^{-1}$. Using the integrated quantity, $\Delta N$, avoids the need to measure the derivative of the projected correlation function, which can be a noisy quantity \citep{Eisenstein:2002sx}. The function $g(x)$ diverges at x=1, i.e. at the bin edges. In principle we can regulate g(x) by adding a cut-off as x approaches 1, but we have not done so here as this does not seem to effect our results significantly for the given number density of subhalos. 

For the discussion below, it will be useful to examine Eq.\ \eqref{eqn:rho} to consider how deprojection works.  Projection causes galaxies at large 3D radius $r$ to appear at potentially smaller 2D radius $R$, according to
\begin{equation}
\Sigma(R) = 2\int_R^\infty \rho(r) \frac{r}{\sqrt{r^2-R^2}}dr.
\label{projected}
\end{equation}
The basic principle underlying Abel inversion is that we can use the density observed at large $R$ to correct for the contributions to the integrand in Eq.\ \eqref{projected} from projections at large radius, to infer the true counts at small radius.  The assumption behind this inversion is that no viewing angles are preferred, so that the population of objects from some 3D radius $r$ will (on average) appear similar regardless of whether they project onto large $R$ or small $R$.  This is why Abel inversion is valid only spherically symmetric distributions.  Even though the individual host halos in the sample plotted in Fig.\ \ref{fig:deproject} are not spherically symmetric, because we view them from isotropically distributed viewing angles, the stacked ensemble is almost spherically symmetric, which is why the deprojected density closely matches the true 3D density profile. In practice, however,  optical clusters are not viewed isotropically, but instead are preferentially viewed along the cluster major axes due to the richness selection of the sample in projection.  This anisotropy invalidates Abel inversion, at least formally.  In practice, however, Eq.\ \eqref{eqn:rho} is still expected to provide a reasonably good estimate of the 3D radial distribution of galaxies around clusters.  This is shown in the bottom panels of Fig.\ \ref{fig:anisotropy}, which applies Eq.\ \eqref{eqn:recon} to hosts selected based on 2D projected richness rather than 3D virial mass. For the purpose of this test, we define richness for each cluster as the number of halos or subhalos with $v_{\rm peak} > 150$ km/s found within 1 $h^{-1}$Mpc projected distance of the cluster center (using centers from the 3D halo catalog), and then select clusters with richness between 20 and 40.\footnote{This method of richness selection is a simplification and not similar in detail to RedMaPPer or CAMIRA, that uses overdensities in red-sequence galaxies to identify clusters and assign richness based on an algorithm that ties the aperture size to the richness.} We project over 50 $h^{-1}$Mpc and 250 $h^{-1}$Mpc and remove the contribution from chance projections. As the figure illustrates, the amount of anisotropy expected from this specific richness selection is not expected to seriously bias the reconstructed space density inferred using Abel inversion. 

In principle, we can use deprojection to reconstruct the 3D distribution not only for all galaxies, but also for subsets of galaxies, such as (for example) red galaxies or blue galaxies.  We can measure their stacked profiles separately, and apply Eq.\ \eqref{eqn:recon} to the separate samples, as long as the `red' or `blue' labels do not systematically depend on viewing angle.  Significant reddening from dust could violate the use of deprojection to infer the red fraction profile in 3D, but in massive clusters dust reddening is expected to be small, with mean extinction of order $10^{-3}$ mag across sight lines of 1 Mpc \citep{Chelouche:2007rm}.

Similarly, we naively could imagine using deprojection to infer the line-of-sight (los) velocity distribution of satellites, as a function of 3D radius $r$, given measurements of the los velocities of projected neighbors as a function of 2D radius $R$.  The basic idea would be to label galaxies according to their observed los velocity relative to the cluster centroid, and then apply Eq.\ \eqref{eqn:recon} separately for each velocity bin, to get the 3D distribution of satellites at each specific observed $v_{\rm los}$.  Generalizing Eq.\ \eqref{eqn:recon}, we could write down
\begin{eqnarray}
\label{eqn:vrecon}
\Delta N(v_{\rm los})&=&4\pi\int_{a}^{b}\rho(r, v_{\rm los})r^2 dr\\ \nonumber
&=&\sum_{i}
\begin{cases}
g(R_i/a)-g(R_i/b) & R_i>b\\
g(R_i/a)+1 & a < R_i<b\\
0 & R_i < a
\end{cases} ,
\end{eqnarray}
where now the sum is not over all the galaxies, but instead is restricted to those with $v_{\rm los}$ falling within a specified velocity bin.
However, applying Eq.\ \eqref{eqn:vrecon} to reconstruct the velocity distribution at each 3D $r$ would not be valid, because line-of-sight velocity is not independent of viewing angle.  Unlike color, line-of-sight velocity is not a scalar, but is instead a component of the velocity vector and is not invariant under rotation.  It is easy to see how this invalidates Eq.\ \eqref{eqn:vrecon}: for projected radius $R$, galaxies at $r\approx R$ have $v_{\rm los}\approx v_{\rm tan}$, whereas galaxies at $r\gg R$ have $v_{\rm los}\approx v_{\rm rad}$, where $v_{\rm rad}$ and $v_{\rm tan}$ are the components in the radial and perpendicular directions, respectively.  Therefore, we cannot use galaxies at large $R$ to correct for interlopers at large $r$ projecting onto small $R$, since the line of sight direction points along different directions in those two cases.  It is only when the velocity anisotropy is negligible, so that the distributions of radial and tangential velocities are similar at each radius, that we can try to use Eq.\ \eqref{eqn:vrecon} to deproject the line-of-sight velocity distribution.

In simulated cluster halos, the velocity dispersion of subhalos is known to be somewhat anisotropic \citep{Cuesta2008}.  In the top panels of Figs.\ \ref{fig:deproject} and \ref{fig:anisotropy}, we show the impact of this level of anisotropy on deprojection of the velocity distribution.  The black curves show the actual distribution of line-of-sight velocities for subhalos in the 3D annulus of $r=0.6-0.9 h^{-1}$Mpc, where the line of sight component for each subhalo lies in the tangent plane to its radius vector. The red curves show the result of deprojection using Eq.\ \eqref{eqn:vrecon}.  The two curves are similar, though not identical, due to velocity anisotropy.

Another potential source of anisotropy in spectroscopic samples arises from the procedure used by the SDSS survey to measure galaxy spectra. 
In fiber-fed spectrographs like SDSS, the finite thickness of the fiber plugs introduces a minimal separation between two adjacent plugs.  As a result, a fraction of the targeted sample of photometric galaxies is absent from the spectroscopic catalog, and these fiber collisions occur more frequently in regions of high projected number density.  This is a well known effect which alters the small scale correlation function.  As the number of galaxies absent at a given location in the sky is proportional to the 2D projected density, instead of the 3D space density, this effect is bound to introduce anisotropy along the line of sight.  For example, in dense regions near the center of clusters, where galaxies are closer in projected space, the probability of losing galaxies is higher than it is in cluster outskirts.  In the Abel deprojection method, since we use the density at large $R$ to correct for the contribution of projection at small radii,  we are essentially overcorrecting to obtain the 3d density.  To test how this affects our reconstruction of the line of sight velocity distribution, we try to emulate a similar loss of objects in our simulations. We randomly remove objects from our sample with a probability proportional to the local projected number density.  We find the probability distribution of projected densities in bins of $\ln(r)$ around our sample of cluster halos, we randomly exclude objects in each $R$ bin around the cluster based on this probability distribution, such that the densest bins (corresponding to an angular scale of 55'' at the median redshift of our sample) lose a fraction $f_{cen}$ of all galaxies.

\begin{figure}
\includegraphics[width=1.0\linewidth, trim= 0.4in 0.0in 0.1in 0in,clip]{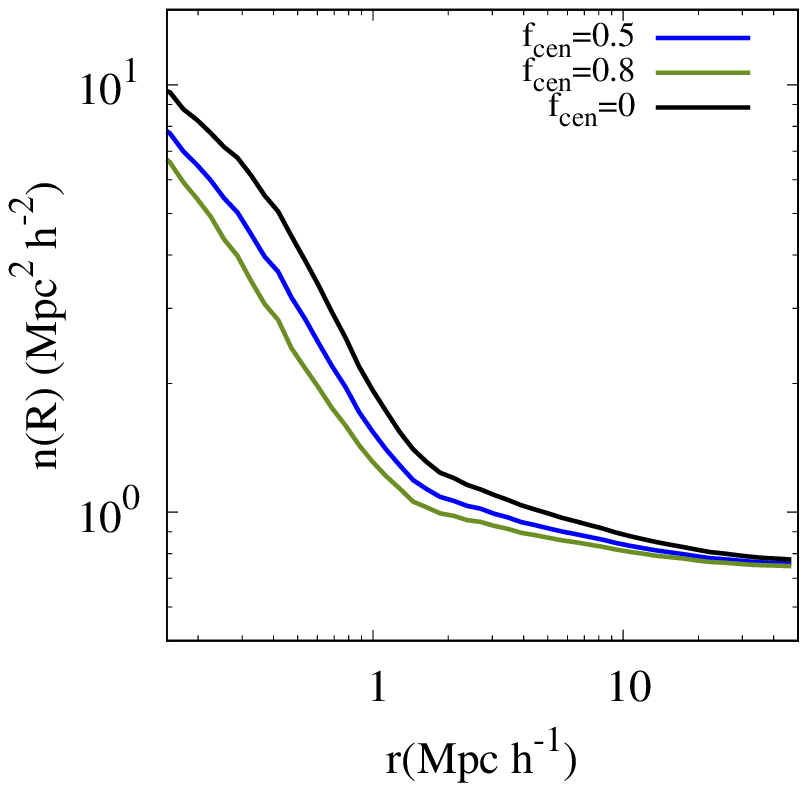}
\includegraphics[width=1.0\linewidth, trim= 0.6in 0.0in  0in 0in,clip]{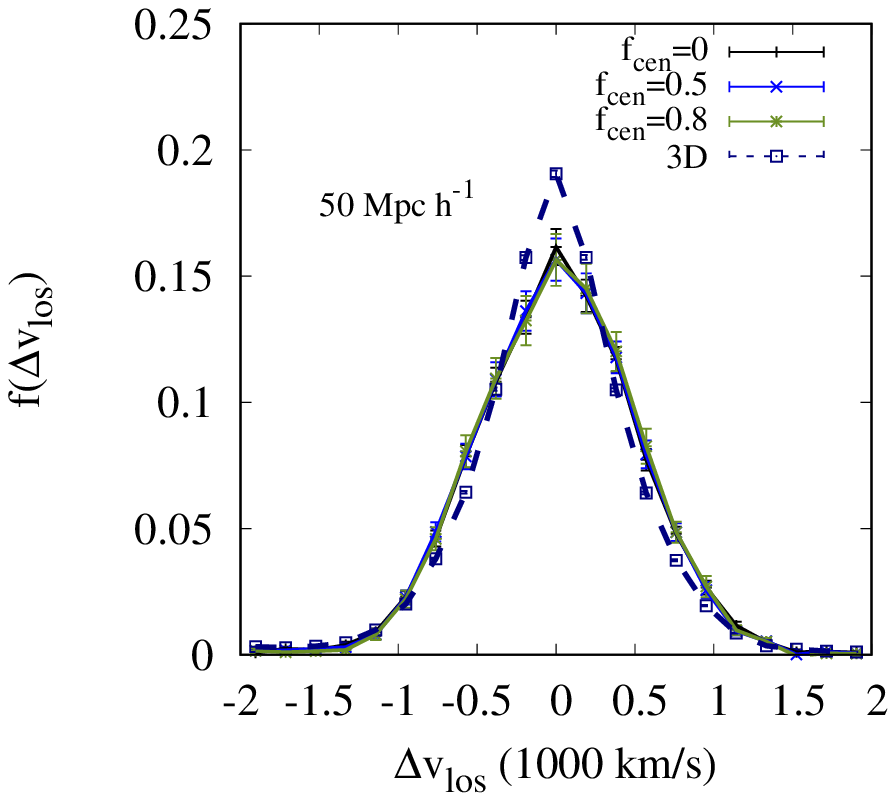}
\caption{Testing fiber collision. Top: Projected density around clusters of mass $1-4\times 10^{14} M_\odot h^{-1}$ when different fractions of objects are dropped from the sample. Bottom: Deprojected LOS velocity histograms in the different cases for $0.6 < r < 0.9$. The black curve in the bottom panel shows the actual 3d distribution LOS velocities in the same bin.} 
\label{fig:fibercollision}
\end{figure}

In figure \ref{fig:fibercollision}, the top panel shows the stacked projected distribution around the cluster host when $f_{cen}=0.5$ and $f_{cen}=0.8$, the bottom panel shows the line of sight velocity histograms for the same cases. While the projected density profile changes considerably, the normalized deprojected line of sight velocity distribution does not change significantly, i.e. no additional bias is added due the loss of galaxies from the sample.

This indicates the extent to which we can use deprojection to infer the true velocity distribution of cluster satellites without interloper contamination.  Anisotropies due to photometric cluster selection or fiber collisions appear to be unimportant, but velocity anisotropy can lead to small biases in the inferred distribution.  This is not necessarily a hindrance, however.  Our aim in this work is to distinguish between different classes of quenching models, using kinematics of blue and red satellites.  If a quenching model predicts that blue and red satellites have identical kinematics, then the two color samples will have identical biases, preserving the similarity of the two reconstructed distributions.  Velocity anisotropy does not appear to limit our ability to detect differences in the kinematics of galaxy samples of different color, as we show below.

\begin{figure*}
	\includegraphics[width=1\linewidth, trim= 0in 2.06in 0.5in 0.35in,clip]{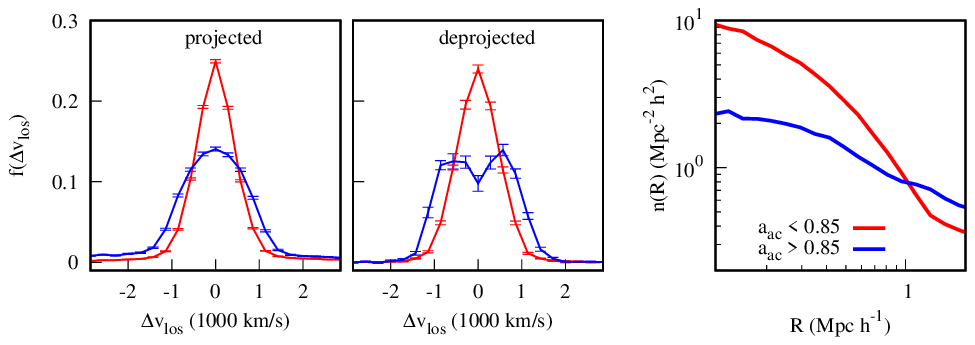}
	\includegraphics[width=1\linewidth, trim= 0in 2.06in 0.5in 0.35in,clip]{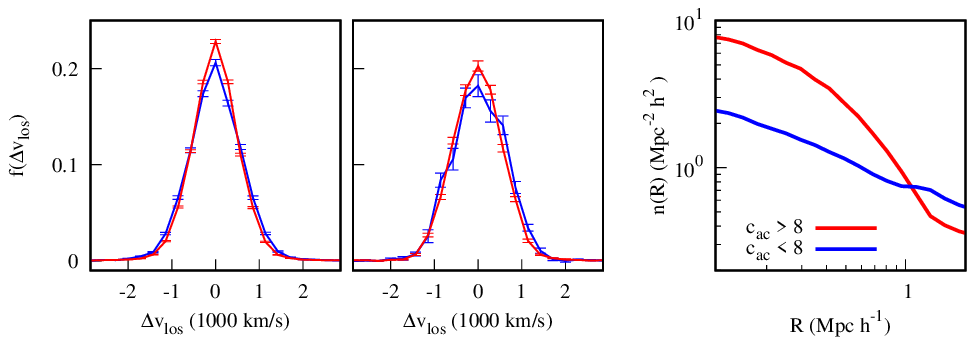}
	\includegraphics[width=1\linewidth, trim= 0in 2.06in 0.5in 0.35in,clip]{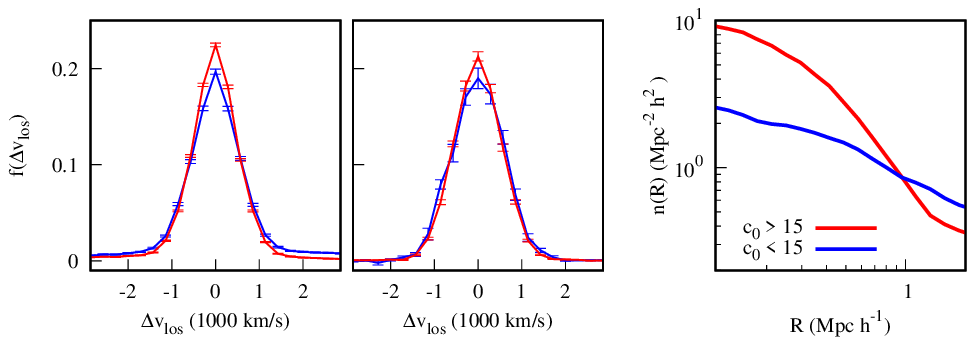}
	\includegraphics[width=1\linewidth, trim= 0in 1.8in 0.5in 0.35in,clip]{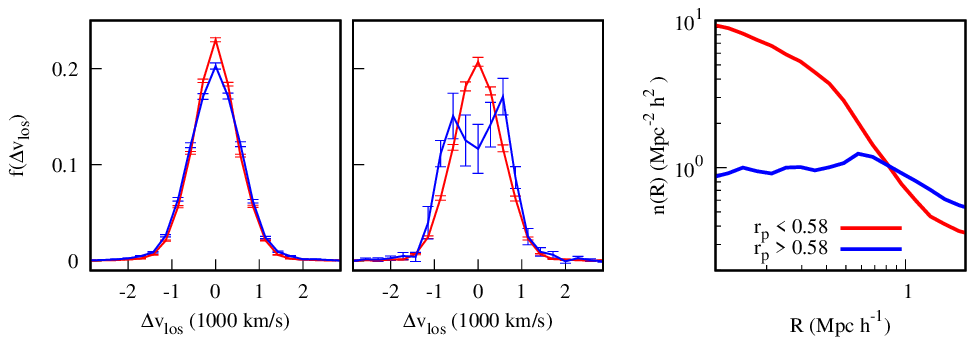}
	\caption{MDPL2 Simulations: The first two columns show the LOS velocity histograms at $0.3$ Mpc $h^{-1}< R <0.7$ Mpc $h^{-1}$ of subhalos with $v_{\rm peak} > 200$ km/s around host halos with mass $10^{14} M_\odot h^{-1} < M < 2\times 10^{14} M_\odot h^{-1}$. The first column shows the projected histograms and the second column shows the deprojected histograms. The last column shows the projected number density of the subhalos around their host halos. The different rows each correspond to various criteria for splitting the subhalos into red and blue. In the top row, subhalos have been split into red and blue based on the time at which they were accreted into any host halo. In the second row subhalos have been split based on their concentration at accretion, in the third row they have been split based on their current concentration, and in the last row they are split based on their pericenters, i.e., based on whether or not they have ever had radii $r < r_p$ within their hosts. \label{fig:sims_all}}
\end{figure*}

To illustrate this, we construct several different possible mock galaxy catalogs from the MDPL2 subhalo catalogs, assigning red or blue colors using various different schemes.  For example, in age-matching mocks \citep{Hearin:2013foa}, galaxy color is determined by the subhalo's age or concentration at accretion $c_{\rm acc}$.  We could similarly assign color using other subhalo properties, such as time of accretion $a_{\rm acc}$.  In Fig.\ \ref{fig:sims_all}, we show velocity histograms for mock galaxies whose `colors' are assigned according to various different subhalo properties.  Note that when constructing projected catalogs, we assign colors for satellites (objects within the hosts' splashback radii) according to the properties described in the figure legend, but for projected interlopers, we assign random colors using a background red fraction of 0.6.  

The top row shows results when we split subhalos according to their scale factor at first accretion into a host.  In this case, the two sets of subhalos exhibit quite different kinematic behavior. Late-accreting satellites with scale factor of accretion, $a_{\rm acc}>0.85$, that have been in their hosts for $\sim2.2$ Gyrs or less,  have a LOS velocity distribution that is much broader than that of the satellites that were accreted early.  Their velocity distribution also shows a double peaked feature that is consistent with orbits with a high angular momentum, in contrast to the early accreters, whose orbits indicate low angular momentum. 
This behavior is consistent with the fact that with time, as the halo radius grows, substructure or particles that are accreted tend to have larger impact parameters and therefore larger angular momenta, creating this signature in the LOS velocity profile.  In contrast, when we split satellites according to their concentration, there is little difference between blue and red kinematics.  Other subhalo parameters can produce other effects on the subhalo distributions.  For example, when subhalos are split based on their pericenters, i.e.\  all subhalos that have pericenters within some $r_p$ are assigned red colors, while those that have larger pericenters (or are on infall and have not crossed $r_p$) were assigned random colors matching the background fraction, we find that ``blue'' subhalos show a slightly broader velocity dispersion but very different radial profile. 

Depending on the subhalo property used to define the mock color, we can find either quite similar velocity distributions for red and blue satellites, or quite different velocity distributions.  Interestingly, this is true even for mocks producing similar spatial distributions of satellite colors.  For example, the first and third rows of Fig.\ \ref{fig:sims_all} have similar number density profiles (compare right panels), but have completely different velocity distributions (compare middle panels).  This result may seem counterintuitive: for example, from the Jeans equation we would expect the velocity distribution to be related to the spatial distribution of subhalos, so it may seem surprising that mocks with similar spatial distributions can have such different velocity distributions.  (Note that the Jeans equations do not necessarily apply to satellites, since their number is not conserved.)  This result demonstrates that satellite kinematics does indeed provide information about satellite properties that is distinct from the information provided by the spatial distribution of satellites.

\begin{figure}
	\includegraphics[width=1.45\linewidth, trim= 1.3in 0in 0in 0.0in,clip]{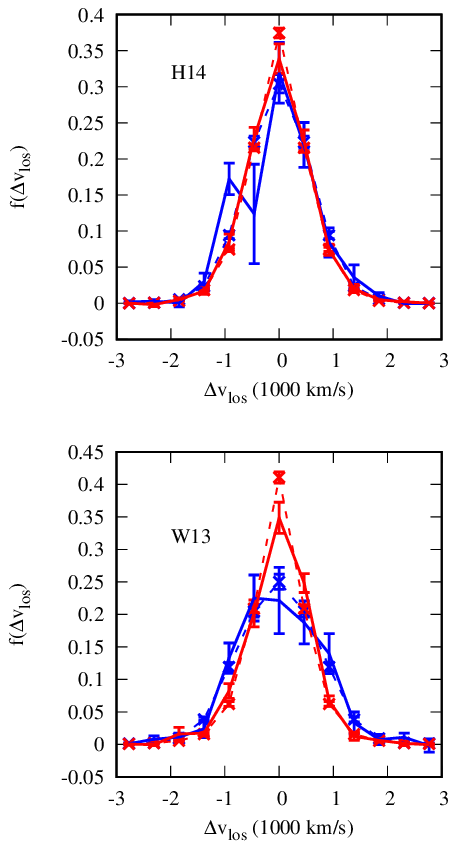}
	\includegraphics[width=1\linewidth, trim=0.2in 0.23in 0.0in 0.0in,clip]{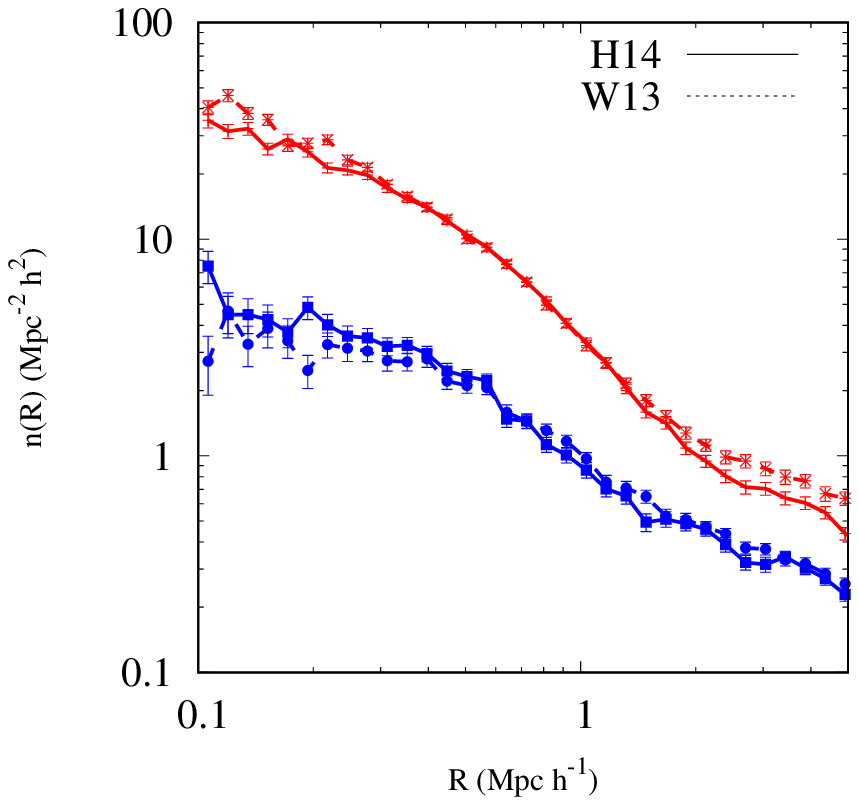}	
	\caption{Comparison between the mock catalogs with age-matching \citep{Hearin:2013foa} and intra-cluster quenching \citep{Wetzel:2012nn}: Deprojected velocity histograms around $10^{14}$ $M_\odot~h^{-1}$ clusters in galaxy mocks between $0.3 < R < 0.7$ Mpc $h^{-1}$. The bottom panel shows the comparison between the spatial clustering of blue and red galaxies in the two mocks. (The dashed lines in the upper two panels show the actual 3D distribution of the subhalos.) 
\label{fig:hearin_wetzel}}
\end{figure}

One consequence of Fig.\ \ref{fig:sims_all} is that satellite kinematics can constrain models for satellite quenching in clusters.  As discussed above, different quenching models relate galaxy colors to different aspects of dark matter subhalos.  In the original age-matching models of \citet{Hearin:2013foa}, for example, galaxy colors are related to $c_{\rm acc}$\footnote{In particular Hearin et al. assign colors to subhalos based on the parameter $z_\textrm{starve}$ which is described as the epoch at which a galaxy is starved of cold gas, $z_{\textrm{starve}}=\textrm{max}[z_\textrm{char}, z_\textrm{form}, z_\textrm{acc}]$. $z_\textrm{char}$ is a characteristic time at which a halo reaches $10^{12} M_\odot h^{-1}$, $z_\textrm{acc}$ is the redhshift at which a halo first becomes a subhalo and $z_\textrm{form}$ is the formation redshift or its age;  which is derived from the concentration of the halo for centrals and the concentration at accretion for satellites based on \cite{Wechsler2002}. For most subhalos $z_\textrm{starve}\simeq z_\textrm{form}$.}
  Fig.\ \ref{fig:hearin_wetzel} shows the comparison between these mock galaxy catalogs. The top two panels of Fig.\ \ref{fig:hearin_wetzel} show the LOS velocity distribution of red and blue galaxies around cluster mass halos in the two mocks. In the top panel, the red curve corresponds to galaxies with $(g-r)>0.8-0.03(M_r+20)$ while the blue curve corresponds to the remainder in the Hearin mocks.  In the Wetzel mocks, we assign colors to satellites based on the time since first infall into any host, $t_{\rm first}$ and centrals are assigned colors randomly with equal probability of being red or blue. We find that the spatial clustering agrees well with the Hearin mocks when galaxies that have $t_{\rm first} > 2.5$ Gyrs are assigned red colors and the ones accreted more recently are assigned blue colors (see bottom panel of Fig.\ \ref{fig:hearin_wetzel}). In both models, galaxy stellar mass is assigned to subhalos based on abundance matching.  We include all galaxies with $M_r< -19$ in the age-matching catalogs and $\log(M_{*}/M_{\odot})> 9.9$ in the Wetzel mocks.  As expected from the MDPL2 simulation results shown earlier, the mocks which relate color to infall time produce a strong color-dependence to the velocity distribution, while the mocks which relate color to satellite properties prior to accretion exhibit no significant color dependence to the velocity distribution. 
Having verified that satellite kinematics can indeed distinguish between models that predict nearly identical spatial distributions of satellites, we next examine real satellite kinematics measurements.

\begin{figure*}
	\includegraphics[width=0.353\linewidth, trim= 1.5in 0.1in 1.5in 0in,clip]{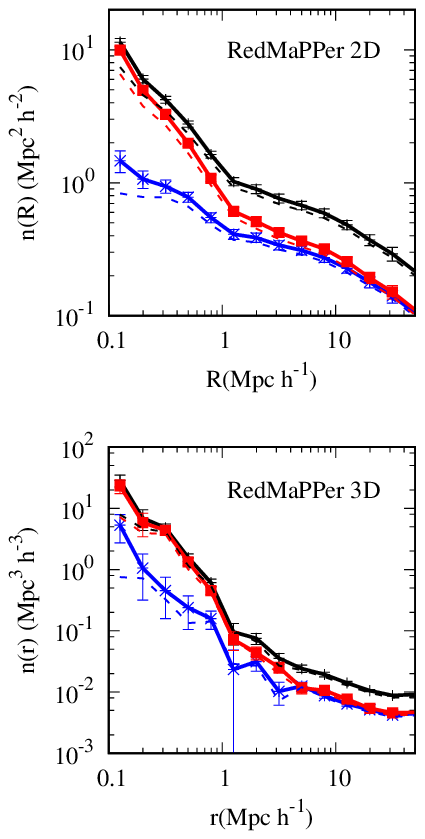}\hfill
	\includegraphics[width=0.3\linewidth, trim= 1.8in 0.1in 1.5in 0in,clip]{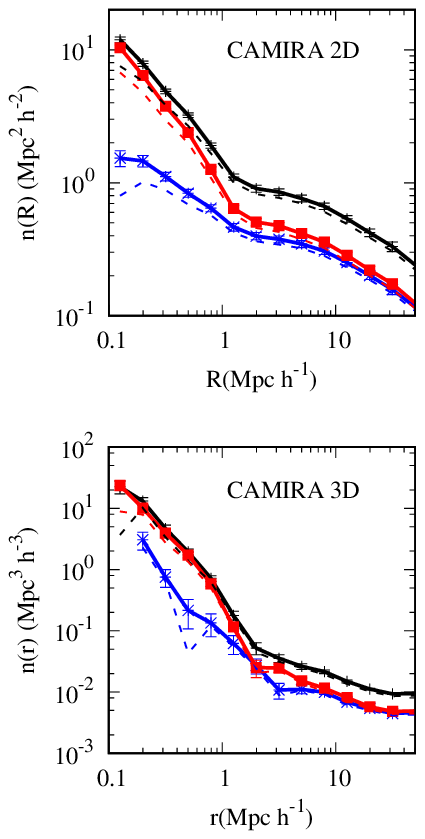}
	\includegraphics[width=0.3\linewidth, trim= 1.8in 0.1in 1.5in 0in,clip]{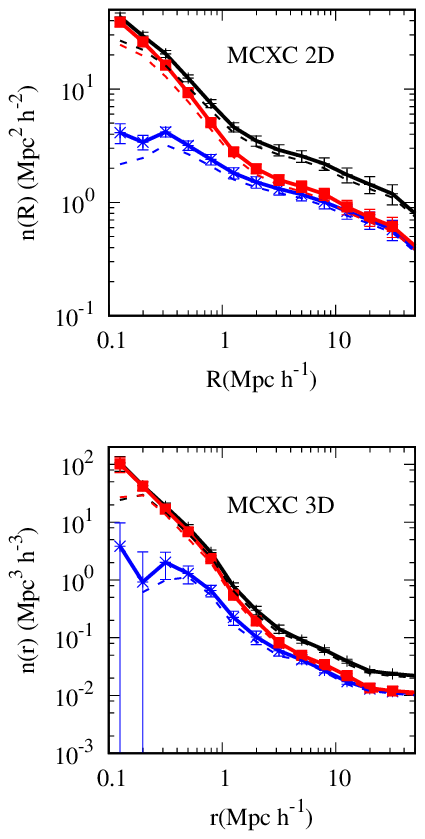}
	\caption{Data from SDSS: The top panel shows the projected number density of spectroscopic galaxies around the clusters RedMaPPer (left), CAMIRA(middle) and MCXC (right). The bottom panel shows the deprojected, 3d number density of galaxies around the same clusters. The black curves correspond to all galaxies, the blue and red curves correspond to blue and red galaxies respectively. The solid lines correspond to the number density measured using the galaxy catalog corrected for fiber-collision, the dashed line is the same but using the the uncorrected galaxy catalog.
\label{fig:dens}}
\end{figure*}

\begin{figure*}	
	\centering
	\includegraphics[width=0.9\linewidth, trim= 0in 0.3in 0in 0.1in,clip]{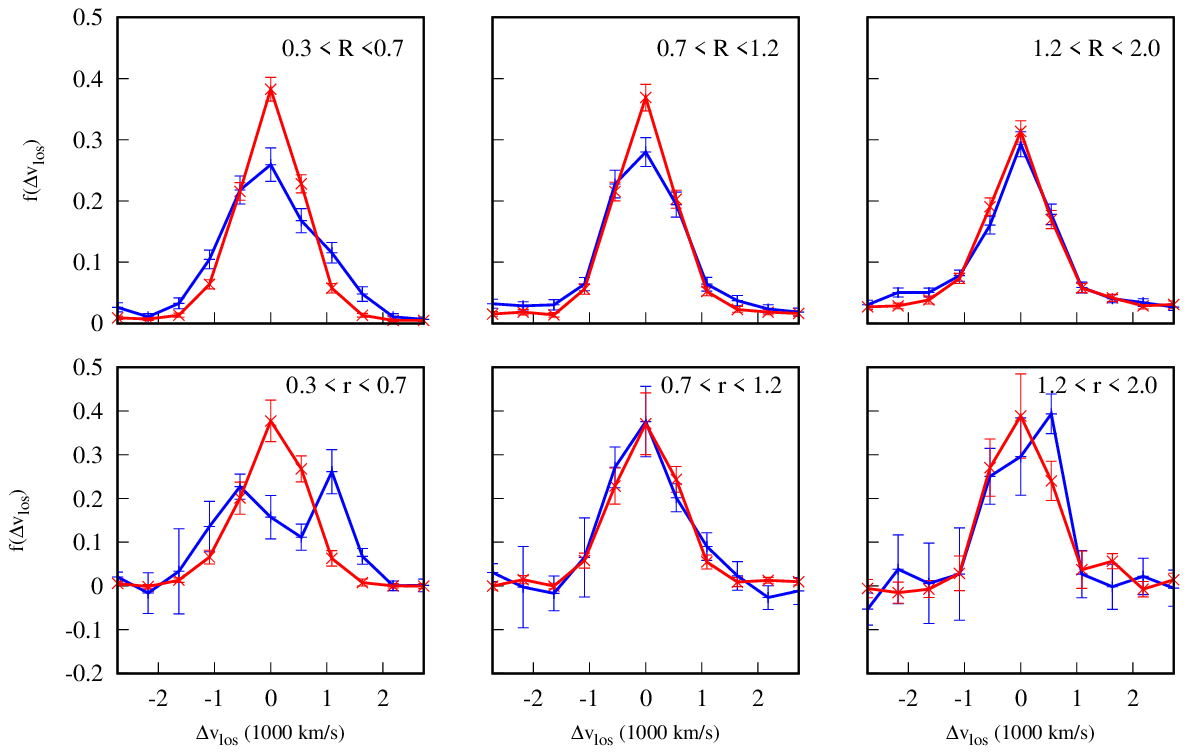}	
	\caption{Data from SDSS: Velocity histograms around RedMaPPer clusters of blue and red galaxies. The top panel shows the projected velocity histograms, and the bottom panel shows the deprojected histograms.\label{fig:obs}}
\end{figure*}

\section{Results from SDSS data}
\label{sec:data}

To determine the kinematics of galaxies in clusters, we require precise spectroscopic redshifts. For this purpose we use the spectroscopic galaxy sample from the eighth data release of Sloan Digital Sky Survey (SDSS). We make three sets of measurements of the distribution of velocities around clusters in catalogs constructed using three different cluster-finding algorithms. We use two different optical samples: the RedMaPPer cluster catalog \citep{Rykoff2014, Rozo2015} derived from SDSS imaging, and the CAMIRA \citep{Oguri14} catalog constructed from Subaru HSC S16A data \citep{Oguri17}.  Apart from the optical catalogs we also use X-ray selected clusters from the MCXC catalog, constructed from the ROSAT all sky survey \citep{Piffaretti:2010my}. We only include clusters that aren't within $1^\circ$ of the boundary of the SDSS spectroscopic footprint. We cross-correlate galaxies and clusters by stacking the number of spectroscopic galaxies as a function of radius around cluster centers.  The SDSS spectroscopic galaxy sample consists of 1200160 galaxies. We select red and blue galaxies based on their $g-r$ galaxy colors, labeling a galaxy as red if $(g-r)>0.8-0.03(M_r+20)$ where $M_r$ is the absolute magnitude in the $r$ band.  Note that
the magnitudes of the galaxies have been k-corrected using the technique defined in \citet{Blanton:2002wv,Blanton:2002jk}. 
The relative velocity between the galaxy and the cluster is given  by, 
\begin{equation}
\Delta v_{los}=\frac{z_g-z_c}{1+z_c},
\end{equation}
where $z_g$ and $z_c$ are the galaxy's redshift and cluster central galaxy's redshift, respectively.

For the optical catalogs, RedMaPPer and CAMIRA, we use the mass-richness relation to select cluster mass halos, and adopt a richness cut $20 < \Lambda < 40$. For both the catalogs we use halos between $0.1 < z < 0.2$, assigning the cluster the measured spectroscopic redshift of the BCG. For RedMaPPer we make an additional cut to exclude galaxies with  $P_{\rm cen}<0.8$. We use 720 clusters in the RedMaPPer sample and 1182 clusters in the CAMIRA sample.  

We also measure the signal in the X-ray selected clusters from the MCXC catalog. This catalog contains clusters which span a mass range from   $11.91< \log_{10}(M500c) > 14.86$. As the number of clusters within the SDSS footprint is not large, we use all the clusters with no mass cut. There are 273 clusters in the X-ray sample with redshift, $0.08 < z < 0.2$. In principle we could also repeat this analysis using cluster samples found spectroscopically rather than photometrically, like the \citet{Yang2007} catalog.  However, one subtlety in using such catalogs is that the cluster selection in velocity space will introduce features in the inferred velocity distribution of nearby galaxies which can confound the measurement of satellite kinematics.\footnote{We thank Frank van den Bosch for explaining this subtlety to us.}  
Rather than attempting to correct for these biases, we have focused only on photometric cluster samples and the X-ray sample.

\begin{figure*}[t]
	\centering	
	\includegraphics[width=0.9\linewidth, trim= 0in 0.3in 0in 0.1in,clip]{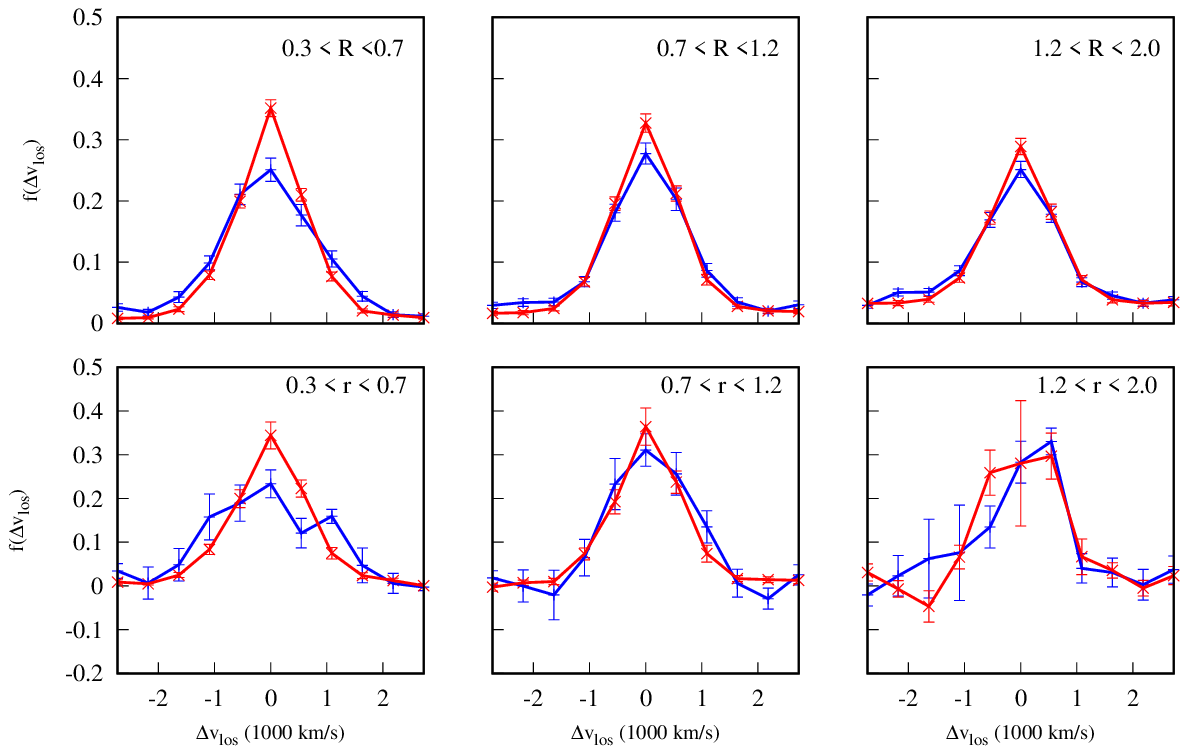}	
	\caption{Data from SDSS: Velocity histograms around CAMIRA clusters \citep{Oguri17} with richness bewteen $20 < \Lambda < 40$. As in Fig.\ \ref{fig:obs}, the top panel shows projected velocity histograms, and the bottom panel shows deprojected histograms.\label{fig:oguri}}
\end{figure*}

\begin{figure*}[t]	
	\centering	
	\includegraphics[width=0.7\linewidth, trim= 0.0in 0.4in 1.5in 0.1in,clip]{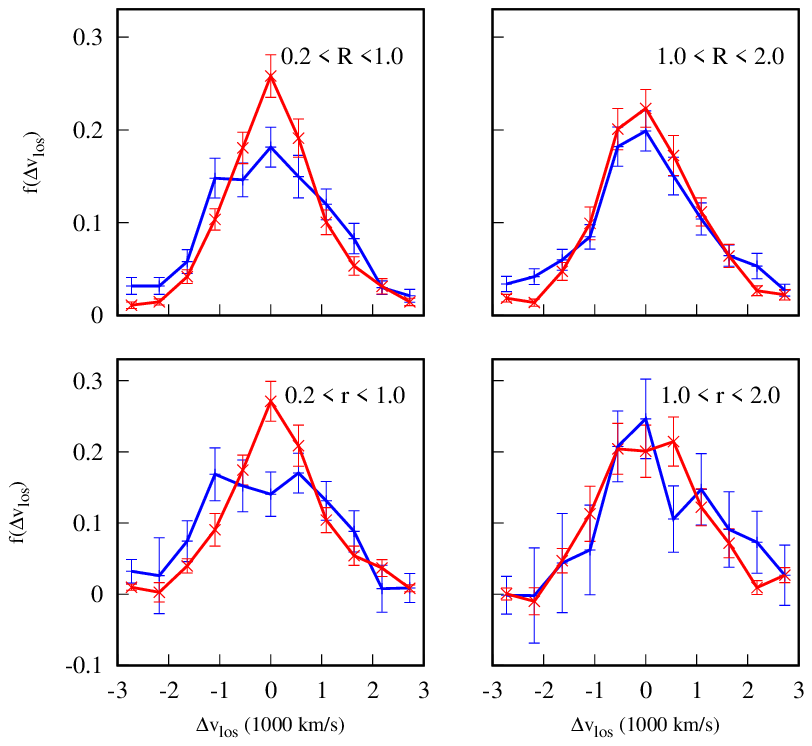}	
	\caption{Data from SDSS: Velocity histograms around X-ray clusters with $\log_{10}(M_{500c})>13.74$. As in Fig.\ \ref{fig:obs}, the top panel shows projected velocity histograms, and the bottom panel shows deprojected histograms.\label{fig:xray}}
\end{figure*}

Fig.\ \ref{fig:dens} shows the projected number density of different population of galaxies around the three cluster samples. The bottom panel shows the deprojected, 3D number density profile of galaxies. The deprojection was done using the method described above for the simulation. The profiles were measured with 15 logarithmic bins between 0.1 and 100 Mpc$h^{-1}$.  The splashback radius acts as the boundary between the virialized and infall regions of the halo. Theoretically this separates the population of galaxies into objects that have completed at least one entire orbit in the halo potential vs.\ those that have never had a pericentric passage and are still in infall. This distinction is important in our context since it serves as an approximate boundary to the cluster's influence.  We measure the splashback radius in 2d using the Savitzky-Golay method from the number density profile of galaxies around clusters. We use the galaxy catalog which has been corrected for fiber-collisions for this measurement. The splashback radius for this sample of RedMaPPer clusters measured using spectroscopic galaxies only is at $0.75\pm 0.03$ Mpc $h^{-1}$, for the cluster in the CAMIRA sample it is at $0.77\pm 0.025$ and the X-ray sample has a splashback at $0.67\pm0.15$.  Although it is possible that the spectroscopic galaxies have a smaller splashback than photometric sample \citep{Adhikari2016} as bright galaxies may be effected by dynamical friction, we only find suggestions of a difference in the splashback radius between spectroscopic and photometric galaxies in the CAMIRA clusters.

Fig.\ \ref{fig:obs} and \ref{fig:oguri} shows the normalized LOS velocity histograms of blue and red galaxies in three different radial bins from the cluster centers in the RedMaPPer and the CAMIRA catalogs. The first radial bin extends from $0.3 < R < 0.7$ Mpc $h^{-1}$, inside the splashback radius. The next two radial bins are between $0.7 < R < 2.0$ Mpc $h^{-1}$, corresponding to the infall region. The red and blue curves correspond to red and blue galaxies as defined previously. The top panel shows the projected LOS velocity histogram. To remove the contribution from galaxies that are projected in the line of sight we deproject our the histograms as was done in the case of the simulations. The error bars on the data points are estimated using the jackknife method. The deprojected velocity distributions for the two galaxy populations are similar in the outskirts of the galaxy clusters. In the innermost radial bin, within the splashback radius, there is a distinct difference between the blue and red population. The dispersion in the blue velocity distribution is larger than that in the red population, consistent with being on orbits with larger angular momentum. Fig.\ \ref{fig:xray} shows the line of sight velocity histograms in projected and deprojected space for the X-ray clusters, here we have extended the inner bin to include objects within $0.2 < r < 1.0$ due to the lower number of objects and a larger uncertainty in the splashback radius. The difference in the LOS velocity histograms between blue and red galaxies persists in the X-ray selected clusters as well. The blue galaxies within these clusters also appear to be on broader LOS velocity distributions. Optical cluster selection methods are known to suffer from projection effects, which leads to confusion in interpretation of their measured richness and related observables \citep{Zu:2016zeu, Busch:2017wnu}. A selection based on richness therefore can introduce anisotropic effects that may bias our deprojection method. While it is unclear how it may generate a signal that depends specifically on the color of the galaxies around them, it is reassuring to see the difference in the LOS velocity histograms in X-ray selected clusters as well. This implies that the difference in the LOS velocity histogram of red and blue galaxies is indeed physical and not an artifact of the cluster selection algorithm. 

\begin{figure*}[t]
	\centering	
	\includegraphics[width=1.0\linewidth, trim= 0in 1.7in 0in 0in,clip]{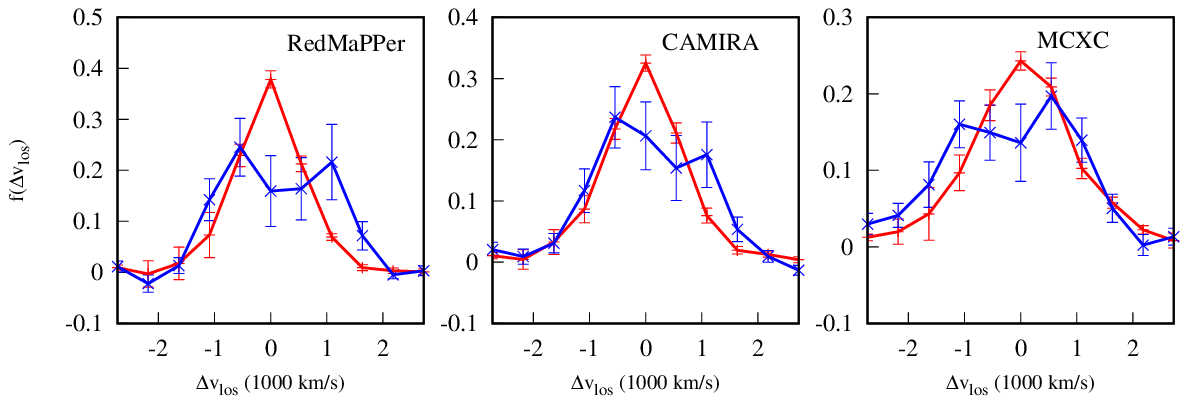}
	\caption{Data from SDSS: Deprojected velocity histograms in the inner radial bins of the three cluster samples split on specific star formation rates. Blue or star-formaing galaxies have log(sSFR)$> -10.9$. The radial bins correspond to $0.3< r < 0.7$ for RedMaPPer and CAMIRA and  $0.2 < r < 1.0$ for MCXC.  \label{fig:ssfr}}
\end{figure*}

To quantify the conjecture that the two populations of red and blue galaxies have been drawn from different distribution functions we use a procedure similar to the Kolmogorov-Smirnov test. This is a model-independent or non-parametric method to test the null hypothesis that two populations have been drawn from the same underlying distribution functions.  
We use the deprojected distributions of LOS velocities to evaluate the cumulative distribution function (CDF) for the red and blue galaxies, and determine the maximum difference between the two CDFs. To quantify how probable such a difference is if the red and blue are drawn from the same underlying population we assign random colors to the galaxies in each radial bin, holding fixed the red fraction of that bin. We generate Monte-Carlo realizations of the randomized LOS distributions and evaluate the maximum difference in the red and blue CDFs for each case. We evaluate how often the maximum difference is equal or larger than the difference in the observed CDFs.  We find that for the innermost radial bin between $0.3 < r < 0.7$ the deprojected distribution of red and blue galaxies are different from each other at greater than $99.99\%$ confidence for the RedMaPPer clusters and $98.7\%$ confidence in the Oguri clusters. For the X-ray clusters, the confidence at which the blue and red profiles are different from each other between $0.2 < r < 1.0$ is $97\%$.  Our method of estimating the significance level of the data from relative measurements of the maximum difference is independent of the number of LOS velocity bins used.  Note that we could also in principle use the projected histograms to quantify the difference between the red and blue samples, since the plots in these figures show significant differences in the inner radial bin even without deprojection.  However, one complication in doing so is that the projected histograms do not vanish at high relative velocity $|\Delta v_{\rm los}|$, which could spuriously lead our test to conclude that the histograms for the two color bins are different.  Since the deprojected histograms all vanish at large $|\Delta v_{\rm los}|$, this particular difficulty can be avoided.

These measurements rely on galaxies in the SDSS spectroscopic catalog, which suffer incompleteness due to fiber collisions, as discussed above.  Our tests using $N$-body simulations give us confidence that fiber collisions do not significantly impact on the recovered velocity distributions.  Nonetheless, we have also empirically tested for the effect of fiber collisions by making use of a catalog that attempts to fill in galaxies that were excluded by collisions, by assigning those galaxies the redshifts of their nearest spectroscopic neighbor.  Even with this extreme assumption, we find no significant changes in the recovered velocity histograms. 

Apart from using the $g-r$ color as a metric for quiescence we also test if the differences in the velocity distributions persist using other spectroscopic metrics like the specific star-formation rate, which is less sensitive to other effects, like dust attenuation and galaxy inclination, than g - r color. We find the similar differences  between the quiescent and star-forming populations. Fig. \ref{fig:ssfr} shows the deprojected velocity histograms in the innermost bins of each cluster sample for galaxies split based on their total sSFRs as described in \citet{Brinchmann:2003db}.

\section{Discussion}

In this paper we have shown that the kinematic behaviors of blue and red satellite galaxies are significantly different from each other within RedMaPPer clusters, inside the splashback radius. At larger radii, in the infall region of clusters, red and blue galaxies exhibit velocity distributions which are indistinguishable from each other.  However inside the cluster boundary, the blue galaxies appear to be on orbits with higher angular momentum. For the blue galaxies in the bin $0.3 < R < 0.7$ there is also mild evidence for bimodality in the histogram, suggestive of tangential motion.

Comparing with simulations, the difference in the red and blue galaxy distributions  closely resembles the difference in the LOS histograms of subhalos classified into different colors based on their accretion time $a_{\rm acc}$. In contrast, simulated satellite kinematics do not seem to strongly correlate with other subhalo properties we examined, like concentration at accretion.  
The significant difference between blue and red satellite kinematics in SDSS RedMaPPer clusters suggests that, like the simulated subhalos, these two samples of galaxies may have accreted onto their hosts at significantly different times on average.  In particular, this argues that red satellites typically entered their host at least one crossing time ago, whereas blue satellites typically entered their hosts relatively recently. This behavior might be expected to arise in environmental quenching models, and it will be interesting to see if it can also be explained in age-matching models as well.  It is important to note that we have used the original catalogs of \citet{Hearin:2013foa}, and though we find a significant tension between the behavior in that catalog and the behavior exhibited by SDSS satellites, it is by no means clear that this potential discrepancy cannot be alleviated by minor modifications to the model.  As an example of this, note that the original age-matching catalogs of \citet{Hearin:2013foa} predicted host halo masses for massive blue centrals that were significantly larger than the masses determined from galaxy-galaxy lensing \citep{Zu2016}, but this discrepancy can be alleviated with only minor adjustments to the age-matching model.  It will be interesting to see if similarly minor modifications to the age-matching model can allow it to match observed satellite kinematics.

While age-matching and intracluster-quenching are two seemingly different ways to account for galaxy quenching, note that the age-matching model of \citet{Hearin:2013foa} does not preclude environmental dependence in the quenching of star-formation. Their age indicator, which in most cases is the concentration of the halo at infall, naturally takes into account the environmental effects around a massive host. Halos in the neighbourhood of massive objects like clusters eventually stop accreting mass due to the tidal field of the massive halo nearby, rendering their density profiles highly concentrated. Our results suggest that, although this environmental effect of reducing the mass accretion on to halos can reproduce the spatial distribution of galaxies as a function of color, simultaneously explaining the velocity distribution may also require dependence of quenching on the nature of orbits of the galaxies within the cluster. It is also interesting to consider this result in light of previous results from SDSS RedMaPPer clusters suggesting that some red satellites entered their hosts recently, based on the persistence of satellite-satellite clustering within clusters \citep{Fang2016}, while at least some blue satellites remain blue after a full crossing time within clusters, based on the presence of a splashback feature in their spatial distribution around clusters \citep{Zuercher_inprep}.  One possible explanation for the latter result, based on the kinematical measurements shown in Fig.\ \ref{fig:obs},  is that the surviving blue satellites may preferentially be on grazing orbits with high angular momentum that avoid the central regions of clusters where astrophysical quenching processes should be most effective.  Taken together, these various results paint a complex picture of satellite quenching, possibly suggestive of contributions from multiple distinct processes. 

It should be straightforward to extend this work considerably, using upcoming spectroscopic datasets.  Upcoming surveys like DESI or WFIRST can not only improve the sample size relative to the sample used here, but can also extend these measurements to lower mass hosts and to higher redshifts.  The same method can also be applied to existing galaxy samples selected using properties other than color.  For example, \cite{OrsiAngulo} argued that emission-line galaxies (ELGs) in clusters are expected to have accreted onto their hosts recently and should be observed on their first infall.  If that is true, then their kinematics would be expected to resemble those of the blue satellites shown in Fig. \ref{fig:obs}.

\section*{Acknowledgments}

We thank Dominik Zuercher for providing the visually inspected MCXC catalog of clusters. We thank Eric Baxter, Shivam Pandey, Arka Banerjee, Bhuvnesh Jain, Risa Wechsler and Frank van den Bosch, for helpful discussions. 
This work was supported by NASA under grant HST-AR-14291.001-A from the Space Telescope Science Institute, which is operated by the Association of Universities for Research in Astronomy, Inc., under NASA contract NAS 5-26555.  
Research at Perimeter Institute is supported by the Government of
Canada through Industry Canada and by the Province
of Ontario through the Ministry of Research \& Innovation.
SM is supported by grants-in-aid from the Japanese Society for Promotion of Science (16H01089) and JST CREST Grant number JPMJCR1414. AW was supported by NASA through grants HST-GO-14734 and HST-AR-15057 from STScI. The CosmoSim database used in this paper is a service by the Leibniz-Institute for Astrophysics Potsdam (AIP). The MultiDark database was developed in cooperation with the Spanish MultiDark Consolider Project CSD2009-00064. The MultiDark-Planck (MDPL2)  simulation suite have been performed in the Supermuc supercomputer at LRZ using time granted by PRACE.




\bibliography{velocity}

\end{document}